# SIFM: A network architecture for seamless flow mobility between LTE and WiFi networks – Analysis and Testbed Implementation

Dhathri R. Purohith and Krishna M. Sivalingam, *Fellow, IEEE*

*Abstract*—This paper deals with cellular (e.g. LTE) networks that selectively offload the mobile data traffic onto WiFi (IEEE 802.11) networks to improve network performance. Several architectures that are proposed based on the IETF Proxy Mobile IPv6 (PMIPv6) framework to support seamless data offloading lacks flow-level mobility support and have a single point of failure. Recently, IETF has proposed extensions to PMIPv6 to support flow-mobility, in which the mobility decisions are done at the Packet Gateway (PGW). This adds complexity at the edge of the LTE core network. We propose the Seamless Internetwork Flow Mobility (SIFM) architecture that overcomes these drawbacks and provides seamless flow-mobility support using concepts of Software Defined Networking (SDN). The SDN paradigm decouples the control and data plane, leading to a centralized network intelligence and state. The SIFM architecture utilizes this aspect of SDN and moves the mobility decisions to a centralized Flow Controller (FC). This provides a global network view while making mobility decisions and also reduces the complexity at the PGW. We implement and evaluate both basic PMIPv6 and the SIFM architectures by incorporating salient LTE and WiFi network features in the ns-3 simulator. Performance experiments validate that seamless mobility is achieved. Also, the SIFM architecture shows an improved network performance when compared to the base PMIPv6 architecture. A proof-of-concept prototype of the SIFM architecture has been implemented on an experimental testbed. The LTE network is emulated by integrating USRP B210x with the OpenLTE eNodeB and OpenLTE EPC. The WiFi network is emulated using *hostapd* and *dnsmasq* daemons running on Ubuntu 12.04. An off-the-shelf LG G2 mobile phone running Android 4.2.2 is used as the user equipment. We demonstrate seamless mobility between the LTE network and the WiFi network with the help of ICMP ping and a TCP chat application.

*Index Terms*—WiFi networks, LTE networks, Inter-RAT Flow mobility, Software Defined Networking, OpenFlow, PMIPv6.

## I. INTRODUCTION

The trend in traffic generation pattern of mobile devices has shifted from text-only data to audio/video data that require high bandwidth. To deal with this, Internet Service Providers (ISP) tend to offload some traffic onto other networks to improve performance. IEEE 802.11 (WiFi), a cost effective and widely deployed wireless technology, is the most popular network for traffic offload. Many ISPs such as AT&T and networking companies like Cisco and Qualcomm have studied architectures to offload the 3G/4G traffic to the WiFi [1], [2]. In data offloading, maintaining the user sessions and offloading of selective flows provides the best user experience in addition to balancing the load on the networks.

The main challenge faced in maintaining the user sessions across networks is that connecting to a different network changes the IP address of the user, resulting in loss of the IP session. The reason for this is, originally Internet Protocols were designed such that the IP address was used to define both the identifier and the location of an entity at the same time in order to reduce the size of the headers. As a result, TCP/IP depends on retaining the same IP address at both endpoints after the movement in order to maintain the session. Therefore, in order to provide seamless user session mobility, the locator and the identifier properties of the IP must be decoupled, which is achieved by IP mobility. Hence, to facilitate seamless data offload, we require the network to efficiently support IP mobility.

IP mobility provides an option to move either all or no flows of a given user across the networks. To better balance the load, we need an option to move selective flows of a given user, i.e. to enable *flow mobility*. Flow mobility provides the user with the flexibility of choosing the most suitable network for a given application. For example, consider a user on a heavily loaded LTE network who is watching a video and also downloading a file. Since video traffic is considered as flexible real-time traffic due to buffering available at the user end, it can be offloaded on to WiFi network. The file download is served with a higher priority on the LTE network itself. The movement of video traffic to the WiFi network frees up more resources on the LTE network for the HTTP data session. This level of freedom in choosing the flows to be moved across the networks cannot be provided only by IP mobility. Thus, flow mobility becomes important in achieving a better distribution of load on the networks along with providing better quality of experience to the users.

There exist several protocol standards for providing IP mobility, namely, Mobile IP (MIP) [3], Dual Stack Mobile IP (DSMIPv6) [4] and Proxy Mobile IPv6 (PMIPv6) [5]. Proxy Mobile IPv6 is a Network-Based Localized Mobility Management Solution (NetLMM) [6]. MIP and DSMIPv6 are host-based protocols, where the user initiates the mobility and thereby requiring significant changes in the mobile node. In PMIPv6, all the mobility related implementations are done

The authors are with the Department of Computer Science and Engineering, Indian Institute of Technology Madras, Chennai India and
India-UK Advanced Technology Centre of Excellence in Next Generation Networks, Systems and Services (IU-ATC)
Email: prdhathri@gmail.com, skrishnam@iitm.ac.in, krishna.sivalingam@gmail.com

at the network and do not require many changes in the mobile node. The 3GPP standard TS 23.261 v12.0.0 proposes a DSMIPv6 based interface for IP flow mobility (IFOM) and seamless Wireless Local Area Network (WLAN) offload [7]. The 3GPP standard TS 24.327 v12.0.0 describes General Packet Radio System (GPRS) and WLAN inter-networking aspects [8]. The 3GPP TS standard 23.402 v13.1.0 proposes architecture enhancements for non-3GPP access [9]. The specification defines interfaces to support network based mobility using PMIPv6. Recently, IETF has proposed extensions for PMIPv6 to support flow mobility [10].

In this paper, we propose a new mobility architecture named Seamless Internetwork Flow Mobility (SIFM), that overcomes the drawbacks of existing architectures by utilizing the concepts of PMIPv6 and Software Defined Networking (SDN) [11]. The SDN architecture is based on decoupling the data plane from the control plane. It introduces two components, namely the controller and the switch. The controller and the switch communicate using the OpenFlow protocol [12]. When a switch receives a packet it has never seen before, it forwards it to the controller. The controller takes the routing decision, and instructs the switch on how to forward similar packets by adding entries in the switch's flow table.

The proposed SIFM architecture exploits the advances in the field of SDNs to handle mobility related control signalling. We define a Flow Controller (FC) similar to an OpenFlow controller [12]. The FC only carries out the mobility related functionality. The Packet Network Gateway (PGW) in the LTE network and the Wireless Access Gateway (WAG) in the WiFi network act as Mobility Agents (MA). They are OpenFlow-hybrid switches that carry out mobility related signalling on behalf of the User Equipment (UE) [12]. They follow the instructions of the FC when a mobile node moves from an LTE network to a WiFi network in order to provide seamless transition.

In the current work, the SIFM architecture and the basic PMIPv6 architecture have been implemented specifically for studying data offloading between the LTE and the WiFi networks. These architectures can be used for traffic offloading between any two access technologies such as 2G and 3G by implementing the functionalities required for seamless mobility at the corresponding entities defined in the respective standards. For the SIFM architecture, the entities that perform the functionality of the MA must be OpenFlow compliant in order to communicate with the FC. Performance studies show that with the best possible (scenario dependent) offload value, both SIFM and PMIPv6 architectures improve performance when offloading the data compared to the no offload scenario. We also show that selective offloading helps in achieving better performance gain by considering a simple scenario and using static flow table rules.

The proposed architecture has been implemented in an experimental testbed consisting of two off-the-shelf LG mobile phones, a software defined radio based LTE base station, and a WiFi access point. The testbed is used to test basic networking functionality between the mobile nodes and other nodes in the network.

The remainder of the paper is organized as follows. Sec-tion II presents the necessary background and related work on mobility protocols. Section III presents the proposed Seamless Internetwork Flow Mobility (SIFM) architecture. Section IV presents the simulation based performance study of the proposed architecture and comparison to the existing PMIPv6 architecture. Section V presents the details of the prototype testbed implementation, and Section VI the testbed based experimental results. Section VII presents the conclusions.

## II. RELATED WORK

This section presents the related work on inter-network mobility protocols.

### A. Network Architecture

The wireless network considered in this paper consists of two types of networks: cellular service provided by a 3G or 4G (LTE) network and access service provided a WiFi network.

The LTE network consists of several LTE evolved NodeB (eNodeB or basestation) nodes that serve the network's cells. The LTE network architecture consists of two major subsystems: the Evolved UMTS Terrestrial Radio Access Network (E-UTRAN) system and the Evolved Packet Core (EPC). The E-UTRAN forms the wireless part of the LTE network and handles the radio communications between the User Equipment (UE) and the EPC. The Evolved Packet Core (EPC) communicates with packet data networks in the outside world such as the internet, private corporate networks or the IP multimedia subsystem.

The EPC consists of three components: (i) *Mobility Management Entity* (MME); (ii) *Serving Gateway* (SGW) and (iii) *Packet Data Network Gateway* (PGW). The MME is mainly responsible for controlling the LTE access network. It is responsible for the activation and de-activation of bearers on behalf of a UE. It also performs tracking and paging procedures for UEs in idle mode and selects the UE's Serving Gateway during the initial attach and handover. The Serving Gateway (S-GW) is responsible for forwarding the end-user's data packets from the eNodeB to the Packet Data Network gateway (P-GW). It serves as the local anchor point during the inter-eNodeB handover.

The Packet Data Network gateway (P-GW) provides connectivity to external PDNs for the UEs. It assigns IP addresses to UEs and may also perform firewall functions such as deep packet inspection and packet filtering on per-user basis. It also performs service level gating control and rate enforcement through rate policing and shaping. One UE can be connected more than one P-GW if it needs to access more than one PDN.

The WiFi network consists of IEEE 802.11 protocol based access points (APs). The APs operate using unlicensed spectrum in the 2.4 Ghz and 5 Ghz bands. The WiFi network is based on collision-based medium access and hence can suffer from collisions when the number of network users is large.

The advantage of the LTE network is its better coverage due to cell ranges that are of the order of a 1-3 Km. In comparison, the WiFi network ranges are typically around 200-300 m. On the other hand, the LTE network capacity per cell sector is of the order of 70 Mbps, with higher capacities supported by the LTE-Advanced standard. The WiFi network's aggregate capacity varies from few hundred Mbps to as high as 1 Gbps based on standards including IEEE802.11ac and IEEE 802.11n.

The current generation user equipment (UE) mobile phones (mostly smartphones) have both 3G/4G cellular and WiFi interfaces. When a UE that is connected to the cellular network comes within the range of a WiFi network, then the UE's data sessions are transferred to WiFi, for two reasons: higher bandwidth and lower (often, no) cost. Thus, there is a need for seamless mobility protocol standards that can transfer a UE's ongoing and new connections from LTE to WiFi and back, as necessary. This aspect is the focus of this paper.

The discussion above presents 3G/4G and WiFi networks. However, this is applicable for any other combination of access protocols that might be developed in the future.

### B. Existing Mobility Support Protocols

The existing standards that define mobility support protocols are presented next.

*DSIMPv6:* The 3GPP Release 8 [7] standard has proposed an architecture for seamless mobility between 3G and WiFi networks based on the Dual-Stack Mobile IPv6 (DSMIPv6) standard [4]. The solution does not require any support from WLAN accesses. The changes are required only at the UE and the PGW. The Home Agent functionality as defined in [3] can be implemented in the PGW or as a stand alone box. The S2c interface is defined between the UE and the PGW for all the DSMIPv6 related communications. Cisco and Qualcomm have proposed architectures based on DSMIPv6 for mobility between the 3GPP and the non-3GPP (WiFi) networks [1], [2]. This approach requires significant changes at the UE since UE initiates the binding related communications, which is the major drawback of the approach.

In UE-initiated procedures, the mobile nodes must signal themselves to the network when their location changes and must update routing states in the Foreign Agent, in the local Home Agent, or in both. This requires changes to the UE stack, and also raises the problem of complex security configurations to authenticate the signalling exchanges and modifications of routing states. The signalling messages also increase the load on the wireless radio interface.

*PMIPv6:* The Internet Engineering Task Force (IETF) has defined a network-based local mobility management protocol, where local IP mobility is handled without any involvement from the mobile node. As a part of the first phase of efforts in this working group, Proxy Mobile IPv6 (PMIPv6), a network-based local mobility management protocol was developed [5]. In network-based mobility, only the entities in the backbone network participate in the mobility related signalling thus reducing the overhead on the wireless radio interface. The changes required at the UE are minimal and the network architectural changes at the backbone are independent of the UE changes unlike UE-initiated procedures in which network changes are tightly coupled with the UE and require changes at the UE side.

PMIPv6 mainly defines two components to achieve network-based local mobility management namely the Local

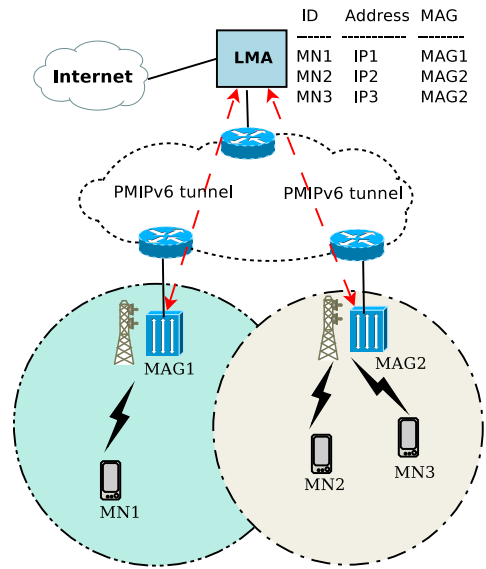

Fig. 1: PMIPv6 architecture.

Mobility Anchor (LMA) and the Mobile Access Gateway (MAG). The MAG performs the mobility related signalling on behalf of the mobile nodes. The LMA keeps track of the users and issues the same IP address to the users when they move across different networks. The LMA and the MAG communicate over the tunnel that is established between the two. Fig. 1 shows the PMIPv6 architecture.

The 3GPP Release 8 standard defines the S2a interface based on the PMIPv6 architecture between the PGW and the Wireless Access Gateway (WAG) for mobility between the 3GPP networks (e.g., LTE) and the non-3GPP networks (e.g., WiFi) [9]. Cisco has proposed an architecture for mobility between the 3GPP and the non-3GPP networks based on PMIPv6 [1].

PMIPv6 solves the problem of user session mobility but does not provide the flexibility to move selected flows of a user. Various architectures adding flow mobility support to PMIPv6 have been proposed in [13] and [14]. Recently, IETF has proposed extensions to PMIPv6 to support flow mobility. In all these architectures, the decision about which flow to move has to be done at the LMA which is generally implemented at PGW. The algorithms that determine optimal movement of flows are quite complex and running them at PGW significantly increases its complexity. The proposed SIFM architecture, exploits the features of SDN by separating the control plane and moving it to a centralized flow controller for mobility related signalling. All the decisions regarding the movement of a flow is done at the Flow Controller instead of PGW, there by reducing the complexity of the PGW. Flow Controller has an overall global view of both the networks, which is an added advantage to make flow mobility decisions at the Flow Controller.

Mechanisms that can be implemented on mobile phones to seamlessly move flows between networks have also been studied extensively. Two such techniques named 'wait-n-migrate' and 'resumption agent' have been proposed in [15]. In 'wait-n-migrate', new flows are established on the new network, but preexisting flows remain on the old network until a specific time-out value set by the policy. 'Resumption agent' resumes a flow from the place it was disrupted, which is transparent to applications. A client based solution that makes switching decisions based on the policies set by the user on his device is proposed in [16]. Qualcomm has proposed a connectivity engine that dynamically determines the characteristics of an unplanned WiFi network on the mobile device [17]. This works together with Access Network Discovery and Selection Function (ANDSF) and IFOM on the network side to provide seamless flow mobility [18], [7]. All these are UE initiated procedures and have drawbacks similar to DSMIPv6 based protocols as discussed before.

A solution for IP flow mobility that can be integrated with either DSMIPv6 or PMIPv6 is proposed in [19]. It introduces a policy routing architecture that is used to signal routing rules between the UE and the network infrastructure. A unified protocol stack that includes all the original functions of both LTE and WLAN systems is proposed in [20]. The solution proposed in [20] consists of a Converged Base Station (CBS) that integrates different Radio Access Technologies (RAT) at the MAC layer. This solution is very complex and requires complete change to the existing infrastructure. The tight coupling between the two RATs poses problems of scalability and restricts independent development of different components in future. An SDN based RAN architecture that provides higher programmability and an easier vertical handover process is proposed in [21]. However, this poses scalability issues. To overcome these problems, we propose a new architecture called SIFM that takes the best of both PMIPv6 and SDN based architectures.

## III. SEAMLESS INTERNETWORK FLOW MOBILITY (SIFM)

This section presents the design, the components and the working of the proposed Seamless Internetwork Flow Mobility (SIFM) architecture.

In this paper, we define a network *flow* as having five attributes: the source and the destination IP address, the source and the destination port and the transport protocol used. A set of packets with common attribute values within a given time period is considered as one flow. The definition of flows can be extended as needed.

### A. Architectural Design

Fig. 2 represents the SIFM architecture for LTE and WiFi networks. The main components of SIFM are: (i) a Flow Controller (FC); (ii) one or more Mobility Agents (MA); and (iii) multiple User Equipment (UE) nodes. In the example shown, the EPC Packet Gateway (PGW) and the Wireless Access Gateway (WAG) act as the MAs. They connect the UE to the Internet and also communicate with the FC. Based on the flow instructions given by the FC, the mobile data flow either takes the path along the LTE network or is offloaded through the tunnel between the PGW and the WAG to reach the UE through the WiFi network.

In the SIFM architecture, we choose to implement the new functionality of mobility using a the centralized controller concepts of Software Defined Networks (SDN). Since the Flow Controller is centralized, it has the complete view of both LTE and WiFi networks. Hence, better algorithms can be developed to dynamically move the flows between LTE and WiFi networks based on the current network conditions, user's charging profile, priorities etc. The communication protocol used for mobility control messages is also based on the OpenFlow standard [12].

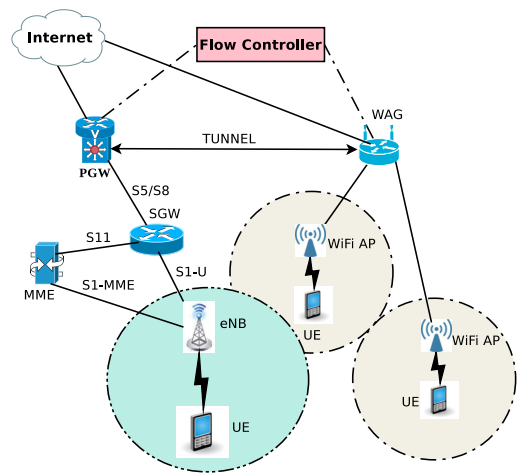

Fig. 2: Proposed SIFM architecture for LTE and WiFi networks.

The problems related to scalability observed in SDN due to centralized controller could be solved by using redundant controllers or having a distributed controller environment [22]. The SIFM architecture is easy to integrate since it does not require major changes to the existing architecture, unlike PMIPv6. This can serve as an intermediate step to move cellular networks towards SDN based architectures.

*1) Flow Controller:* The FC is a new component defined in the SIFM architecture and has to be added to the current LTE EPC in order to facilitate flow mobility. The FC works similar to an OpenFlow controller but only performs the *flow mobility related* actions in this architecture [12]. The Flow Controller (FC) is responsible for assigning flows to the Mobile Agents (MAs), defined in the next section. It is aware of all the MAs to which the UE is connected to and is active on. Based on the information obtained from the MAs, the FC sets up the flow rules and communicates the same to the MAs. When a UE moves from one network to another, the FC instructs the MAs involved to create a tunnel through which active connections are transmitted to the UE without any disruption.

*2) Mobility Agents:* The Mobility Agent (MA) is a router that provides Internet services to the UE and is responsible for detecting the movement of the UE between the access technologies (e.g., LTE and WiFi). The MA's functionality is similar to that of the MAG in PMIPv6 [5]. In the OpenFlow context, it behaves similar to an OpenFlow-hybrid switch [12]. Whenever a UE comes within a MA's access network, the MA assigns an IP address to the UE and informs the FC about its binding. The MA gets the flow information from the FC and forwards the data packets based on the obtained information. The MA can provide more information about the UE's binding such as access technology, link bandwidth information, packet and port statistics to the FC. Based on these information, the FC can run several algorithms to assign flows between the MAs. Only control messages are exchanged between the MAs and the FC. There is no actual data transfer between the two. A messaging protocol, similar to OpenFlow, is used to communicate between the FC and the MAs. In the current infrastructure, functionality of the MA is added at the PGW for the LTE and at the WAG for the WiFi network.

*3) User Equipment:* The UE is a user or a mobile node (MN) that requests for the service. User, UE and MN mean the same and are used interchangeably in this paper. Since the SIFM architecture supports flow mobility, the UE should be able to receive packets destined to multiple IP addresses at the same time. Also, the IP address cannot be bound to a network interface, in case of the flows which are being moved from LTE to WiFi or vice-versa. To support this, either the UE should support a weak host model or have a logical interface [23], [24]. The logical interface abstracts the underlying physical interfaces called the sub-interfaces and may be attached to multiple access technologies (e.g., LTE and WiFi). Only the logical interface is exposed to the higher layers at the UE and the physical interfaces are hidden. The Transmit/Receive functions of the logical interface are mapped to the Transmit/Receive services exposed by the sub-interfaces. This mapping is dynamic and any change is not visible to the upper layers of the IP stack.

### B. Data Structures

Two main data structures are defined to support SIFM, as described below: Binding Cache (BC) at the FC and Flow Table (FT) at the MA.

*1) Binding Cache (BC):* The binding cache (BC) is maintained at the FC and contains the information about all the UEs and the MAs to which they are attached to. Every BC entry contains:

- *MN-ID:* Identifies the Mobile Node uniquely.
- *MA-ID:* Identifies the Mobility Agent uniquely.
- *MN-IP:* IP Address of the Mobile Node within MA's network.
- *MA-IP:* IP Address of the Mobility Agent. This is the tunnel address which is used to communicate with other MAs.
- *PORT-ID:* Physical/Logical port on which the packets destined to the MN are forwarded at the MA.
- *STATUS:* Status of the UE in a MA's network.

*2) Flow Table (FT):* The Flow Table (FT) is maintained at the MAs and reflects the flow decisions taken by the FC. Entries of the FT determine the path taken by the flow received at the MA. Each FT entry contains:

- *Match-fields:* Fields of the packet header to match against the incoming packets. It might be an ingress port, source/destination IP etc.
- *Priority:* Matching precedence of the flow entry.
- *Counters:* Updated when the packets are matched.
- *Instructions:* Actions to be performed by the MA on the packets matched.
- *Timeout:* Idle time before a flow is expired by the switch.

### C. Message Formats

Communication between the FC and the MAs uses messages that comply with the OpenFlow Protocol. Two new message types that follow the experimenter message type as specified in [12] are introduced. The Flow Modification Message and the Port Status update Message defined in OpenFlow are used to meet the system requirements.

*1) Binding Update:* A Binding Update is sent from an MA to the FC when the MA receives a connection request from the UE. The Binding Update message contains MN-ID, MA-ID, MN-IP, MA-IP, PORT-ID and STATUS. MA-IP is the tunnel IP of the MA which is used to communicate with the other MAs. On receiving the Binding Update message, the FC updates its BC and sends a Binding Acknowledgement.

*2) Binding Acknowledgement:* On receiving the Binding Update, the FC sends a Binding Acknowledgement to the MA to acknowledge the reception of the Binding Update. In addition to this, if the FC already has a BC entry for the UE, it sends the information about the IP of the UE corresponding to the old MA, to the new MA. This is required by the new MA to map the flows corresponding to the old IP, to the new IP.

*3) Flow Modification Message:* A Flow Modification Message is sent from the FC to an MA to inform the MA about all the flow mobility related decisions taken by the FC. The Flow Modification Message mainly contains information about the match fields which is used to match the incoming flow at the MA and the instructions which define the actions to be performed on the matched flows. In addition to this, it also contains auxiliary fields such as priority and time-out values. On receiving the Flow Modification Message, the MA updates its Flow Table.

*4) Port Status Update:* The Port Status Update is sent from an MA to the FC to inform the FC about any change in the UE's port status with respect to the MA. It contains MN-ID, MA-ID, PORT-ID and STATUS. On receiving the Port Status Update, the FC updates the status of the port as per the received BC entry in the appropriate BC entry.

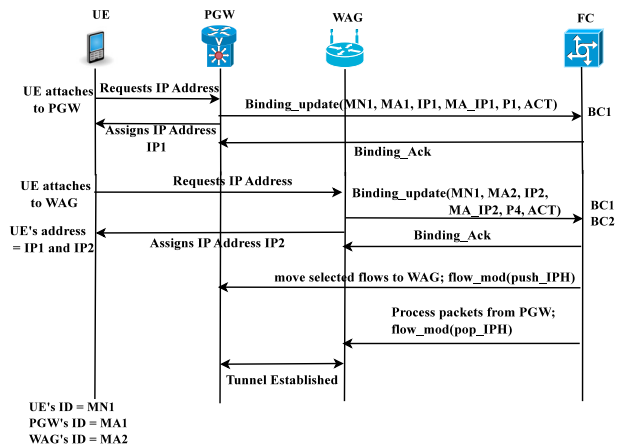

Fig. 3: SIFM architecture operation details.

### D. Protocol Operation

Fig. 3 explains the working of the SIFM architecture with the LTE and the WiFi networks. The steps are detailed below.

1) The UE connects to the LTE network. The PGW receives the connection request, assigns an IP to the UE, sets up default bearers and sends the Binding Update to the FC. The FC responds with a Binding Ack.

2) When the UE comes in the range of a WiFi network, the WAG receives a connection request from the UE. The WAG assigns a new IP to the UE and sends a binding update to the FC. Both the IP addresses are now configured on the UE's logical interface.

3) Since the FC already has an entry corresponding to the UE, it makes a new entry for the WAG and sends a Binding Ack to the WAG. The FC also sends Flow Modification Messages to the PGW and the WAG based on its decision.

4) In case of dual network connectivity ( i.e., UE is in the range of both LTE and WiFi networks), the FC can run several algorithms based on the link bandwidth information, packet and port statistics, etc. and assign flows between the PGW and the WAG based on the algorithm's output. Algorithms should be aimed at balancing the load on both the networks and provide better quality of experience to the user.

5) In case of complete handover (i.e., UE moves out of the LTE network range and connects to WiFi network), the PGW informs the FC about the UE's movement by sending a Port Status Update message. The FC instructs the PGW to move all the existing flows corresponding to the UE over the tunnel. It also instructs the WAG to decapsulate the packets received on the tunnel from the PGW and forward it to the UE on the appropriate port. The new connections from the UE are established over the new network (i.e, WiFi in this case), there by reducing the tunnelling overhead.

### E. Discussion

In the PMIPv6-based Distributed Mobility Management (DMM) approach [25], the functions of LMA (both control and data plane) are split across multiple MAGs and hence distributed. The proposed solution combines the best of both centralized and distributed approaches. The control and the data plane are separated, and the LMA which hosts the control plane is still a centralized entity. However, the data plane is distributed across the MAGs. At a higher level, even though the concept is similar to the partial DMM approach, the proposed work defines and evaluates a more fine-grained architecture.

This paper has considered a domain to be within that of a region served by a given PGW. when a mobile node moves from one PGW to another, it is considered to be moved to a different domain: this case is NOT handled in this paper. And even though PGW covers wide-areas, the controller can be made scalable by using replicated controllers with consistency mechanisms. Any solution to SDN controller scalability can be used in our framework.

## IV. PERFORMANCE STUDY: SIMULATION MODELING

This section explains the simulation modeling, topology and the network parameters used to evaluate the PMIPv6 and the SIFM architectures for LTE and WiFi networks. Comparative analysis of the metrics for both the architectures are presented.

The PMIPv6 based architecture and the SIFM based architecture for LTE-WiFi network mobility have been implemented in the ns-3 network simulator [26] (version 3.20). The ns-3 network simulator does not support IPv6 with LTE. To implement PMIPv6, we have extended the LTE module of the simulator (LENA) to support IPv6. PMIPv6 messages i.e, proxy binding updates and acknowledgements are implemented as per RFC 5213 [5]. LMA is implemented as a different node which is connected to PGW/SGW node and WAG by point-to-point links. Communication between the LMA and the access gateways happen via PMIPv6 tunnel that is established at the start of the simulation.

To implement the SIFM architecture for LTE and WiFi networks, we have added a new module to ns-3 simulator called "openflow-hybrid". This implements the functionalities of the FC, OpenFlow-hybrid switch and the messaging between the FC and the MA. The PGW/SGW and the WAG are modified to support the MA functionalities. To support the logical interface at the UE, a new *UeLogicalNetDevice* class, that abstracts the LTE and the WiFi interfaces at the UE, is implemented.

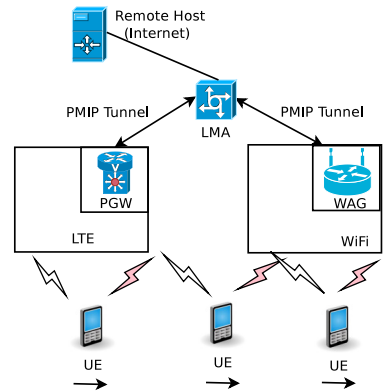

Fig. 4: Network topology for PMIPv6 architecture.

### A. Network Topology and Simulation Parameters

The network topology for PMIPv6 architecture is as shown in Fig. 4. Both the PGW and the WAG (MAGs) are connected to LMA via a point-to-point link and communicate over the PMIPv6 tunnel. The LMA is connected to the Remote Host.

The network topology for the SIFM architecture is as shown in Fig. 5. In the SIFM architecture, both the PGW and the WAG are directly connected to the Internet. The PGW and the WAG (MAs) are connected to the FC via a point-to-point link and a TCP connection is established between the FC and the MAs for all mobility related communications. A simple IP-in-IP tunnel is established between the PGW and the WAG to offload the traffic. In all the simulation experiments, we create the tunnel pro-actively at the start of the simulation. This can also be dynamically created (re-active approach) when there is a need to offload. But this approach is not considered since it might increase the overhead of the handover process.

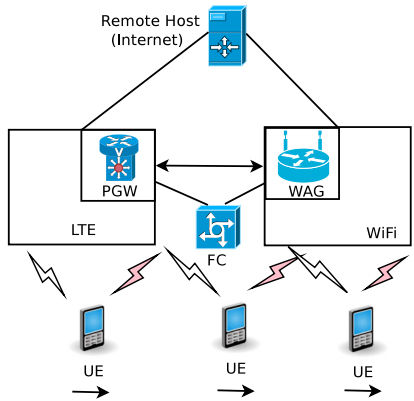

Fig. 5: Network topology for SIFM architecture.

The network parameters used for the simulation are summarized in Table I. The effective capacity of LTE downlink is ≈ 71 Mbps, uplink is ≈ 40 Mbps and that of WiFi network is ≈ 22 Mbps due to signalling overhead. The effective WiFi Bandwidth is actually the total available bandwidth with 802.11a. Even though the theoretical maximum bitrate is 54Mbps, in reality only 22Mbps is achieved, due to effects of inter-frame spacing gaps in the MAC protocol. This has been verified using the simulations. This does not change with the number of users. As the number of users increase, the bandwidth per user decreases; however, the available bandwidth (capacity) is still the same.

Multiple UEs are connected to either the LTE or the WiFi Network. We consider a single eNB to which all the UEs are connected. Both TCP and UDP traffic are running between every UE and the remote host. The details of the traffic are as given in Table I. Some of the UEs are static and others are mobile as determined by the offload value specified in the simulation. The offload value indicates the percentage usage of the available WiFi bandwidth (≈ 22 Mbps). The arrow in Fig. 4 and Fig. 5 indicates the direction of movement of the UEs. Initially, all the UEs are in the range of the LTE Network. The UEs that are mobile will move towards the WiFi network and their flows are offloaded when they get connected to the WiFi network. The speed of the UE is 1 metre/s (1.8 Kmph). A lower speed is considered since WiFi offloading helps only when users are less mobile. In other words, UEs are expected to be connected to WiFi for a sufficient period of time such as at home, at coffee shops, or in the office. When a user is constantly moving at a high speed, offloading induces a ping-pong effect since the user will have to shift between LTE and WiFi networks frequently. The algorithms at the flow-controller are responsible to detect this and not offload sessions for such users. Since, this work does not concern algorithms at the flow-controller, we study only low-mobility (or almost stationery) scenarios. The delay, the throughput and the packet loss values are calculated for different offload percentages by varying the number of UEs. However, for low to moderate speeds, offloading can be applicable.

TABLE I: Simulation Parameters

| Parameter | Value |
|---|---|
| Bandwidth of Link Connecting to Internet | 1 Gbps |
| LTE Downlink Capacity | 100 Mbps |
| LTE Uplink Capacity | 50 Mbps |
| Scheduler Used at LTE | Proportional Fair |
| WiFi Network Capacity | 54 Mbps (802.11 a) |
| No. of users | varied from 10 to 50 |
| Offload value | varied from 0% to 75% of WiFi Bandwidth |
| Traffic at each UE | One UDP app, 1 Mbps (CBR & VBR); One TCP app, 1 Mbps (CBR & VBR) |

### B. Performance Evaluation Results

This section presents a comparative analysis of delay and throughput for the SIFM architecture and the PMIPv6 architecture considering different offload values. The packet loss values are not reported due to space constraints. The offload value indicates the percentage of maximum feasible WiFi bandwidth. For a 54 Mbps link, this is taken to be ≈ 22 Mbps, to account for various inter-frame spacing overhead. This has been obtained using simulations with a single user in the WiFi network, and is taken as an upper limit of the feasible WiFi bandwidth. The experiments are repeated for both Constant Bit-Rate (CBR) and Variable Bit-Rate (VBR) traffic. The packet arrival for VBR traffic follow a Poisson distribution with a mean rate of 1 Mbps and the results are presented with 95% Confidence Interval. The graphs presented shows the results for VBR traffic only. The CBR traffic also shows similar behaviour.

*Delay:* Fig. 6(a) presents the average delay experienced by a UE for different offload values for both the architectures. This includes the average delay of both TCP and UDP flows. It can be seen that up to 30 users, LTE successfully handles the load and the average delay is less than 300ms since the total capacity of the LTE eNB is approx. 71 Mbps and the total downlink traffic is 60 Mbps for 30 users. When the number of UEs is further increased, the LTE network experiences congestion and hence the delay increases for the no-offload scenario. For 40 and 50 users, offloading the data to WiFi network decreases the average delay. The degree of improvement is less for 50 users since the traffic on the network (approx. 100 Mbps) is more than the combined bandwidth available from both LTE and WiFi networks (approx. 93 Mbps). However, when the load on the LTE network is less, higher offload value increases the delay. For example., in 10 users and 75% offload scenario, the delay is higher compared to the no-offload scenario. This increase is because of higher load (75% of the available bandwidth) in the WiFi network despite the bandwidth being available on the LTE network.

TABLE II: Impact of RLC Buffer Size on Delay

| RLC buffer size | Avg. delay (ms) |
|---|---|
| 10 KB | 13,063.30 |
| 100 KB | 9,917.26 |
| 2 MB | 8,880.96 |
| 10 MB | 7,639.10 |

Table II presents the variation of average delay for different

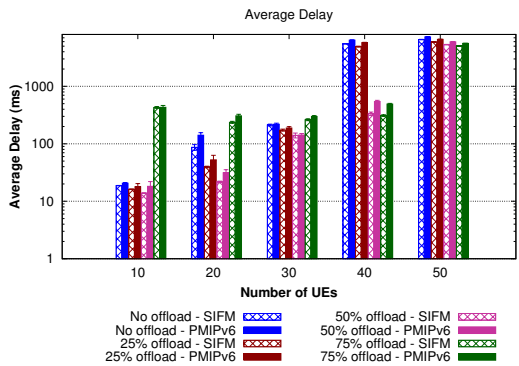

(a) Delay

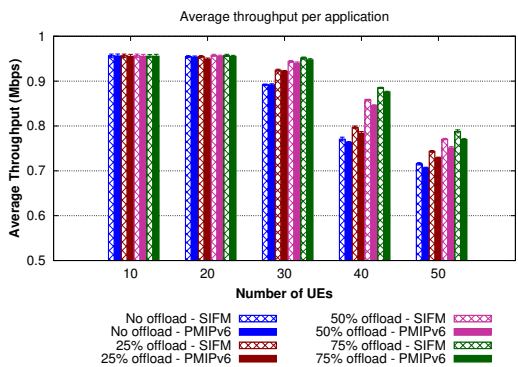

(b) Throughput

Fig. 6: Comparison of SIFM and PMIPv6 architectures.

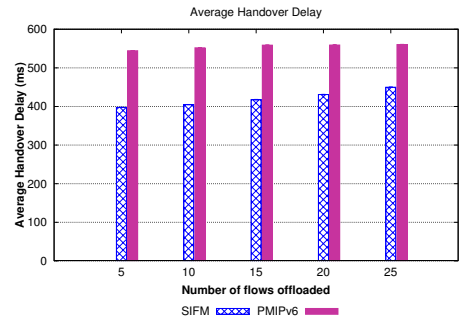

Fig. 7: Handover delay comparison for SIFM and PMIPv6.

Radio Link Control layer (RLC) buffer sizes for 50 users in the LTE network. When the buffer size is small and the network is loaded, the packets are dropped at the RLC layer, which triggers re-transmissions of TCP packets and thus increases the delay. However, when the buffer size is large, TCP does not see any loss of packets, thus re-transmissions are reduced and hence the delay is reduced. However, in order to reduce fragmentation at the lower layers, RLC buffer is kept small. This adversely affects the performance of TCP applications when the network is highly loaded. One other thing to note in this experiment is that, for each of the RLC buffer size, TCP buffer size is kept constant and the change in the delay is only due to RLC retransmissions and hence the delay keeps decreasing with increase in buffer size.

*Throughput:* Fig. 6(b) presents the average per application throughput experienced by a UE for different offload values for both the architectures. It can be seen that the throughput is almost equal to the requested bandwidth (i.e, 1 Mbps/application) up to 30 users. Further increase in the number of users decreases the throughput drastically for the no-offload scenario. Offloading the traffic for higher number of users (40 and 50 users), i.e. when the LTE network is congested, increases the average throughput. In addition, for a high load scenario, increase in the offload value increases the throughput since more of the available WiFi bandwidth is being used.

*Handover delay:* Fig. 7 presents the average handover delay for a flow that is moved between the LTE and the WiFi network. PMIPv6 has a slightly higher handover delay. This is due to the fact that when a UE moves to a different MAG, the IP address configuration takes time since the MAG has to consult the LMA before sending the Router Advertisement.

*UE-level performance:* Fig. 8 present the average delay and throughput at any time instant for an arbitrarily selected user plotted against the simulation time. These graphs are plotted for an arbitrarily selected UE in the 40-user, 50%-offload scenario and show how offloading benefits this user.

Fig. 8(a) presents the average per packet delay at any instant 't' of TCP and UDP applications for an arbitrarily selected LTE user and a user whose flows are offloaded to WiFi. We can see that the delay values go down for both the users when the offloading is done around 1s. Around 14.8s, the offloaded users come back to LTE network from WiFi network. Hence the delay increases again.

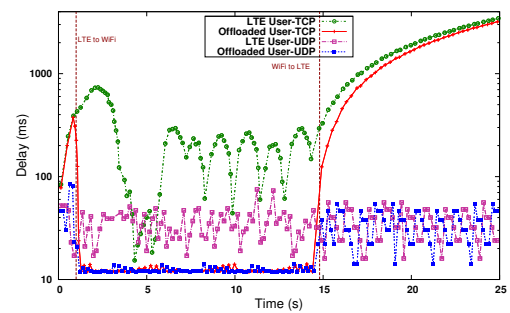

(a) Instantaneous delay

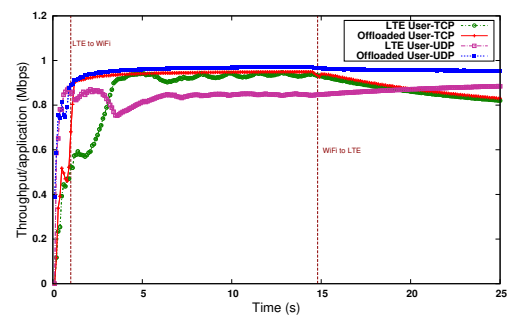

(b) Throughput

Fig. 8: Performance of an arbitrarily selected UE.

Fig. 8(b) presents the average throughput of TCP and UDP applications at any time instant 't' for an arbitrarily selected

LTE user and a user whose flows are offloaded to WiFi. We can see that the throughput for both the applications are higher for the user who is offloaded to WiFi compared to the throughput of the user who stays in LTE network. This is because the offloaded user gets more bandwidth in the WiFi network when compared to the loaded LTE network. The throughput decreases again when the offloaded user moves back to LTE network around 14.8 seconds.

*Flow Mobility:* The main advantage of the SIFM architecture over the PMIPv6 architecture is flow mobility support. Flow mobility is useful when the user is connected to both the networks at the same time. To demonstrate the advantages of flow mobility, we consider the same topology as in Fig. 4 and Fig. 5 but static UEs. All the UEs are connected to both LTE and WiFi networks at the same time. We consider 3 kinds of offloads:

- *Full Mobility*: All the flows of a user is moved.
- *TCP offload*: Only TCP flows of a user is moved.
- *UDP offload*: Only UDP flows of a user is moved.

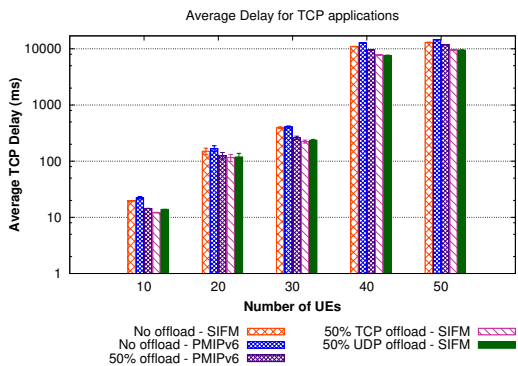

(a) TCP delay

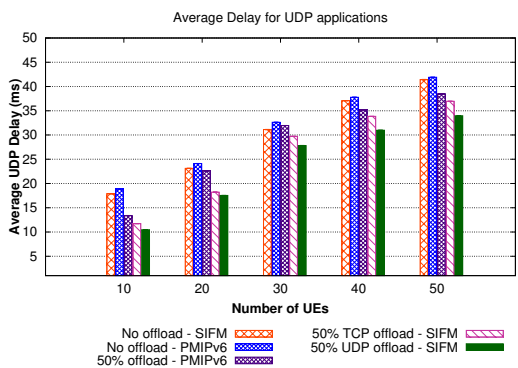

(b) UDP delay

Fig. 9: Delay comparison: with and without flow mobility.

The results are presented only for 50% offload scenario due to space constraints. In case of PMIPv6, there is no selective offload. Hence equal number of TCP and UDP flows which together amounts to 50% of the WiFi bandwidth are offloaded. For TCP offload, only TCP flows equivalent to 50% of the WiFi bandwidth are offloaded and for UDP offload, only UDP flows equivalent to 50% of WiFi bandwidth are offloaded. Full mobility scenario for PMIPv6 is compared with selective offload scenario for the SIFM architecture. The results are shown in Fig. 9 and Fig. 10.

Fig. 9(a) shows that the reduction in the average delay for TCP applications is higher in the selective offload scenario compared to full mobility in PMIPv6. This is demonstrated clearly in the graph for 40 and 50 users. Fig. 9(b) presents a similar result for UDP applications. Fig. 10(a) and Fig. 10(b) show that the applications experience a better throughput in selective offload scenario compared to full mobility scenario.

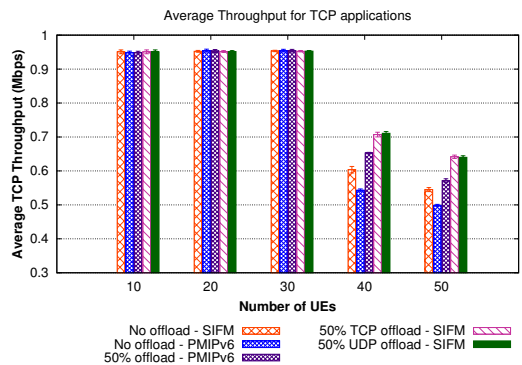

(a) TCP Throughput

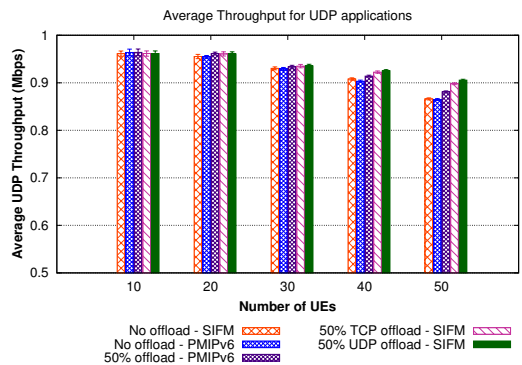

(b) UDP Throughput

Fig. 10: Throughput comparison: with and without flow mobility

|  | Complete Offload | Selective Offload |
|---|---|---|
| **TCP delay** | 7.02% | 22.72% |
| **UDP delay** | 16.32% | 21.39% |
| **TCP packet loss** | 15.79% | 19.17% |
| **UDP packet loss** | 11.76% | 20.56% |
| **TCP throughput** | 2.64% | 3.16% |
| **UDP throughput** | 0.24% | 0.28% |

TABLE III: Average performance improvement of SIFM over PMIPv6 architecture

Table III presents the average performance improvement of the SIFM architecture compared to PMIPv6 architecture (average over all the scenarios mentioned before is considered). The experimental results show that flow mobility provides more flexibility in offloading the traffic and in turn achieves better performance gain. For example, flows can be given priorities and offloaded based on the priorities. In the experiments conducted, if UDP flows are assigned high priority, then the results show that retaining the TCP flows and offloading only UDP flows to WiFi will improve the overall performance of the UDP traffic compared to the full mobility scenario. Similarly, if TCP traffic is given high priority, then offloading only the TCP traffic will improve the performance of the TCP applications. Using algorithms for dynamic flow modification based on user and flow priorities and the current network state, additional performance gain can be achieved.

Only control packets flow between the FC and the MA in the SIFM architecture. The FC is no more a single point of failure because it is only logically centralized and only the mobility related functionality is affected when the FC fails. Since the data and the control plane is separated, failure of the FC does not have any impact on the functionalities of LTE and WiFi as stand alone networks. Only the existing flows pass through the tunnel between the MAs, and the new connections are established over the network to which the UE is currently attached to. Thus, the load on the tunnel is reduced.

## V. Testbed Implementation

This section explains the set-up and implementation of the proof-of-concept prototype of the SIFM architecture.

### A. Testbed Set-up

Fig. 11 shows the topology of the experimental testbed being implemented. The PGW, the WAG and the FC are implemented on Linux systems that are connected to each other over the LAN (at DON Lab, IITM). USRP b210x board is connected to the system running as PGW, which provides the radio interface for the LTE network. WAG is implemented on a Linux system with WLAN interface support. Remote Host is also a Linux system connected to the LAN (at DON Lab, IITM).

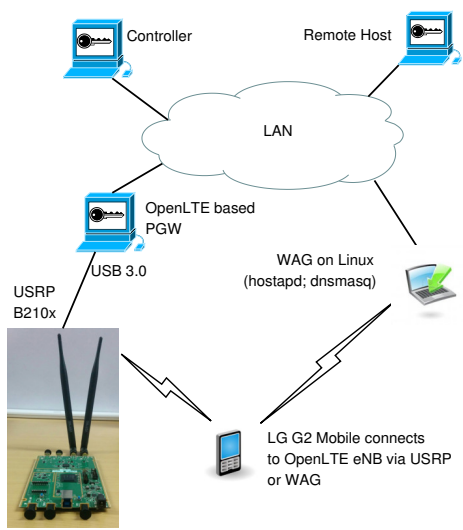

Fig. 11: Testbed Topology.

### B. LTE Network

The LTE network is emulated using an open-source implementation of the 3GPP LTE specifications known as OpenLTE [27]. OpenLTE includes the implementation of LTE eNB with a built-in LTE EPC. It also includes tools for scanning and recording LTE signals based on GNU radio [28]. GNU radio is a software framework that can be used with external RF hardware to create software-defined radios (SDR). OpenLTE currently supports ETTUS radios (USRP B200x and USRP B210x) for the external RF hardware [29]. USRP Hardware Driver (UHD) along with GNU radio acts as an interface between the OpenLTE eNB and the USRP hardware and communicates the RF signals between the two. OpenLTE requires huge amount of processing power and a very low latency since it transmits and receives a radio frame every 1 ms. Any delay in processing results in loss of radio signals. Hence, it is recommended to run OpenLTE on a machine that has high processing capabilities and also to turn off any system processes that can cause delay due to context switching time.

We have extended the existing OpenLTE PGW functionality to support seamless data offloading. A new TCP socket was created to send and receive messages to and from the FC. The Flow Table was implemented at the PGW that defines the routing for the packets received. The Flow Table rules are updated by the Flow Modification messages that are sent by the FC. The uplink path is unaffected by the changes. For the downlink packets, the PGW first checks its Flow Table for a match. If there is no match, then the packets are forwarded to the UE via OpenLTE eNB and the USRP radio. If there is a match, appropriate action is taken, which in the case of offload is to send the packet to UE via the WAG. The offloaded packets are sent to the WAG over a pre-established IP-in-IP tunnel between the PGW node and the WAG node.

### C. WiFi Network

The WiFi network is emulated by creating a WiFi hotspot on a laptop running Ubuntu 12.04. A Linux box with at least one WiFi interface and one Ethernet interface can be turned into a Wireless Access Gateway (WAG) with the help of *hostapd* and *dnsmasq*. *Hostapd* is a user-space program that implements access point functions and authentication servers [30]. The *dnsmasq* software implements the DNS forwarder and DHCP server [31].

To support seamless data offloading, a new "*wificlient*" interface is implemented that communicates with hostapd and the FC. The *wificlient* connects to the control interface of hostapd and listens to the events that are sent by the hostapd. It also implements a TCP socket that is used to send and receive messages from the FC. The *wificlient* interface captures the events such as the connection and the disconnection of the UE to the WiFi network via hostapd control interface and informs the FC about the events by sending Binding Update and Port Status Update messages. It also takes care of routing the incoming packets based on the Flow Table rules that is updated by the FC via Flow Modification Messages.

### D. Flow Controller

The Flow Controller (FC) was implemented as an user-space application that can run on any Linux system. The FC communicates with the OpenLTE PGW and the WAG over a TCP server socket that can handle multiple clients. On receiving Binding Updates and Port Status Updates from the OpenLTE PGW and the WAG, the FC updates its Binding Cache entry and replies with a Binding Ack and a Flow Modification message whenever necessary. In the present implementation, to show the seamless transition, we move the flow over WiFi network whenever a UE comes in contact with the WiFi network. The FC software can be extended to implement algorithms to determine when and what flows could be moved.

### E. User Equipment

An off-the-shelf commercial LG G2 mobile phone running Android 4.2.2 is used as the User Equipment (UE). The OpenLTE's Home Subscriber Station (HSS) database is updated with the mobile's dummy International Mobile Subscriber Identity (IMSI) so that authentication is successful and the UE successfully connects to the LTE network. In order to support seamless mobility, both LTE and WiFi interfaces should be active and capable of receiving packets simultaneously. Android by default allows only one active interface at a time. To support multiple active interfaces, routing functionalities of Android should be modified or a logical interface should be implemented both of which requires a lot of changes in the Android source code. As a work around, we have developed an application that exploits the weak host model supported by Android and use the High Priority (HIPRI) feature. This allows only certain connections to go over the cellular network even when WiFi is on, thereby keeping both the cellular interface and the WiFi interface active at the same time. HIPRI basically lets the programmer have data as high-priority even when WiFi is available.

## VI. Testbed Evaluation

This section presents the results of the experiments conducted on the proof-of-concept testbed. We demonstrate that a TCP connection can be moved seamlessly from the implemented LTE network to the WiFi network. The aim of the prototype was to demonstrate basic functionality and feasibility. Getting the system integration done was quite technically challenging and time-consuming, when compared to running simulation based experiments. Since the OpenLTE package is not very robust enough, we could not conduct detailed performance runs on the testbed at this point and is left for future work.

### A. Single UE communicating with a server on the network

This section presents the results of the experiment in which a UE communicates with the remote TCP server connected to the LAN. Fig. 12(a) shows the OpenLTE eNB running on the Linux machine. OpenLTE loads the required firmware on the USRP B210x board and tunes it to the required frequency to listen to LTE signals. Before starting the OpenLTE eNB, the parameters should be set for the radio. The user information is added to the Home Subscriber Station (HSS) of OpenLTE for authentication. Fig. 12(b) shows the configuration of the

eNodeB and the successful attach of the user to the created OpenLTE network. Fig. 13 presents the screen shots of the connected mobile device. At the right top corner of the UE screen we can see the icon *4g*, denoting the LTE connection. The figure also shows successful ping to a remote machine on the LAN over the established LTE connection.

We demonstrate the seamless mobility feature of the SIFM architecture with the help of a TCP chat application. TCP server runs on a remote machine in the LAN. Fig. 14(a) shows that the UE is connected as a TCP client to the remote TCP server over the established LTE connection. Fig. 14(b) presents the messages that are exchanged between the TCP server and the client over the LTE connection. The red circle at the right top corner of the UE screen shows that only LTE is active at this time.

After a while, the WiFi AP is turned on, and the UE connects to this network too. The red circle at the right top corner of the UE screen in Fig. 14(c) shows that both LTE and WiFi are active at this time. As discussed in Section V-E, we can currently show only the movement of downlink traffic due to the limitations of Android. We require the uplink packets to still go via LTE network in order to keep both the LTE and WiFi interfaces active. The figure presents the messages exchanged between the TCP client and the TCP server over the WiFi network.

Fig. 15 presents the packet traces captured at the PGW and the WAG nodes. This helps establish that the downlink traffic to the UE is sent over the WiFi network. We can see both uplink and downlink packets in the packet trace at the PGW (Fig. 15(a)), where as the packet trace at the WAG (Fig. 15(b)) shows only the downlink packets. Downlink packets are exchanged between the PGW and the WAG over the IP-in-IP tunnel created between the two. The UE receives the downlink packets over the WiFi interface.

### B. Two UEs communicating with each other

This section presents the results of the experiment in which two UEs communicate with each other, after successfully attaching to the created OpenLTE network. Fig. 16(a) shows the 4G connection on both the UEs and the successful ping between the connected UEs over the OpenLTE network. One of the UE runs a TCP client and the other runs a TCP server. The TCP server IP address is 192.168.1.4 and the client IP address is 192.168.1.3. Fig. 16(b) shows the connection between the TCP client and the TCP server. It also presents the messages exchanged between them. The 4G connection at the right top corner of the screen shot validates that the connection is over the LTE network.

After a while, only the client is moved to WiFi network. In the Fig. 16(c), it should be noted that only the client UE is connected to WiFi and the server remains in the LTE network. It can also be observed that the established TCP connection is intact even after the client is moved to WiFi network. The messages from the TCP server are offloaded successfully from LTE network to the WiFi network over the IP-in-IP tunnel created between the PGW and the WAG.

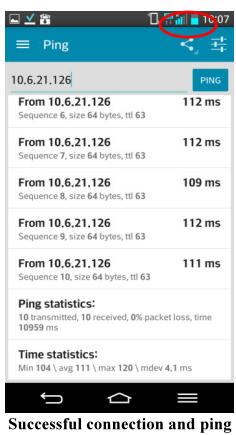

(a) OpenLTE eNB running

(b) UE connected to LTE network

Fig. 12: Screen shots of eNodeB and UE, showing connection establishment messages.

Successful connection and ping

Fig. 13: Mobile connected to the created test network.

*C. Discussion*

In this paper, the solution to data offloading is looked from a network perspective rather than user perspective. However, the decisions about which flows to move, can still consider user preferences. It is assumed that both LTE and the WiFi services are provided by a single operator or there is a Service Level Agreement (SLA) between the two operators in order to move the flows between the LTE and the WiFi networks. With the increase in the use of mobile users, we believe that the LTE service providers like AT&T who also have their own WiFi infrastructure can deploy more and more WiFi access points for data offloading. There can also be an agreement between the existing LTE service providers and the WiFi service providers in which the WiFi providers allow the LTE traffic to be offloaded with the appropriate charging policies. Since all the offloaded traffic go via the PGW, Policy and Charging Functionality (PCRF) at the LTE core network can handle the charging policies. Authentication between the two networks is an important aspect in data offloading which is a problem on its own and is not discussed in this paper. The 3GPP standard TS 24.234 v12.2.0 proposes EAP SIM and EAP AKA based authentication mechanisms that can be used for authentication [32]. The scalability and reliability problem of the Flow Controller has not been dealt in this paper. There have been several research that is going on to build a scalable and reliable SDN controller. Any such solution can be used to have a reliable and scalable Flow Controller (FC).

## VII. Conclusions

In this paper, we present a new architecture called Seamless Internetwork Flow Mobility (SIFM) for the network based

(a) TCP connection between UE and Remote host

(b) Messages sent over LTE network

(c) Messages sent over WiFi network after offload

Fig. 14: Flow Mobility between LTE and WiFi.

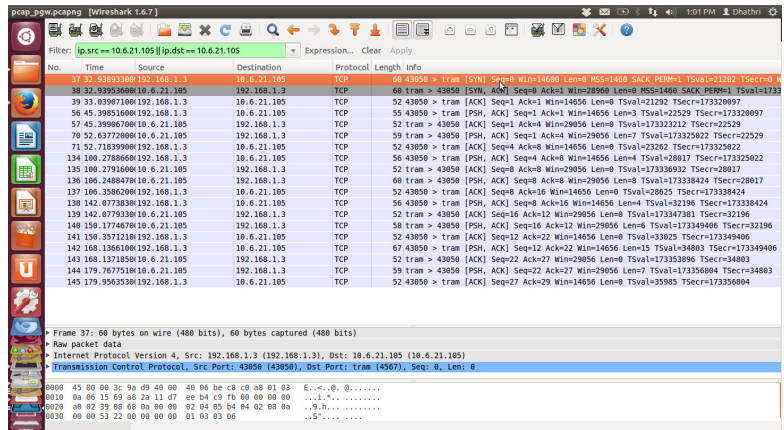

(a) Packet trace at PGW

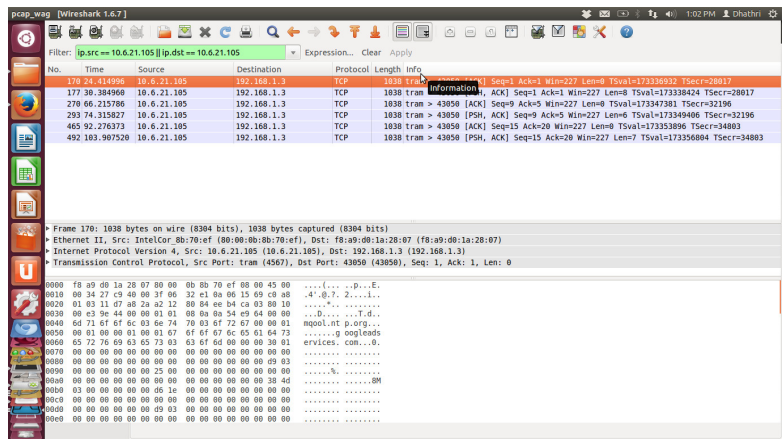

(b) Packet trace at WAG

Fig. 15: Traces captured during WiFi traffic flow.

flow mobility and seamless data offload. The SIFM architecture and the PMIPv6 architecture have been implemented in the ns-3 simulator and evaluated for data offloading between the LTE and the WiFi networks. The evaluation results show that the delay, the throughput and the packet loss values are improved (on an average) by 16.86%, 1.58% and 16.82% respectively for the SIFM architecture compared to the PMIPv6 architecture. It is shown that the flow mobility support provides the flexibility to move selective flows which helps in achieving better performance gain compared to moving all the flows of the user. A proof-of-concept prototype of the SIFM architecture is implemented on an experimental testbed. We demonstrate the seamless movement of a TCP flow between the implemented LTE and WiFi networks to support the working of the SIFM architecture. In future work, mechanisms for determining when to switch from LTE to WiFi and dynamically modifying the flow table rules based on the flow priority and current network state can be studied.

## REFERENCES

[1] "Architecture for Mobile Data Offload over Wi-Fi Access Networks," http://www.cisco.com/c/en/us/solutions/collateral/service-provider/service-provider-wi-fi/white_paper_c11-701018.html, Dec. 2015.

[2] "3G LTE Wifi Offload Framework," http://www.qualcomm.com/media/documents/3g-lte-wifi-offload-framework, Dec. 2015.

[3] D. Johnson, C. Perkins, and J. Arkko, "RFC 3775: Mobility Support in IPv6," Jun. 2004.

[4] H. Soliman, "RFC 5555: Mobile IPv6 Support for Dual Stack Hosts and Routers," Jun. 2009.

[5] S. Gundavelli, K. Leung, V. Devarapalli, K. Chowdhury, and B. Patil, "RFC 5213: Proxy Mobile IPv6," Aug. 2008.

[6] J. Kempf, "RFC 4830: Problem Statement for Network-Based Localized Mobility Management (NETLMM)," Apr. 2011.

[7] "3GPP TS 23.261 : IP flow mobility and seamless Wireless Local Area Network (WLAN) offload, (Rel. 10)," http://www.3gpp.org/DynaReport/23261.htm, Dec. 2015.

[8] "3GPP TS 24.327 : Mobility between 3GPP Wireless Local Area Network (WLAN) interworking (I-WLAN) and 3GPP systems; General Packet Radio System (GPRS) and 3GPP I-WLAN aspects, (Rel. 10)," http://www.3gpp.org/DynaReport/24327.htm, Dec. 2015.

[9] "3GPP TS 23.402 : Architecture enhancements for non-3GPP accesses, (Rel. 10)," http://www.3gpp.org/DynaReport/23402.htm, Dec. 2015.

[10] C. Bernardos, "RFC 7864: Proxy Mobile IPv6 Extensions to Support Flow Mobility," May 2016.

[11] W. Stallings, *Foundations of Modern Networking: SDN, NFV, QoE, IoT, and Cloud*. Addison Wesley, 2015.

[12] "OpenFlow Specification," https://www.opennetworking.org/images/stories/downloads/sdn-resources/onf-specifications/openflow/openflow-spec-v1.4.0.pdf, Dec. 2015.

[13] H.-Y. Choi, S.-G. Min, and Y.-H. Han, "PMIPv6-based Flow Mobility Simulation in NS-3," in *5th International Conference on Innovative Mobile and Internet Services in Ubiquitous Computing (IMIS)*, Jun. 2011, pp. 475–480.

[14] T. M. Trung, Y.-H. Han, H.-Y. Choi, and H. Y. Geun, "A design of network-based flow mobility based on Proxy Mobile IPv6," in *3rd IEEE International Workshop on Mobility Management in the Networks of the Future World*, Apr. 2011, pp. 373–378.

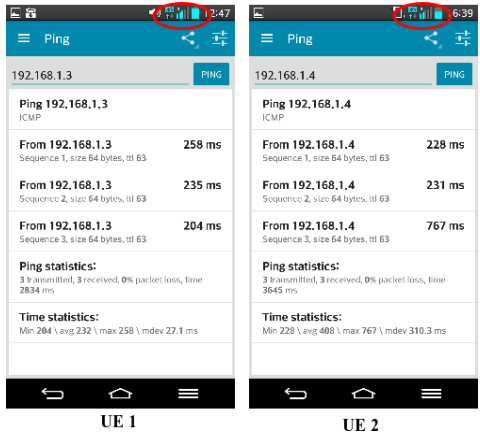

(a) Ping between 2 UEs

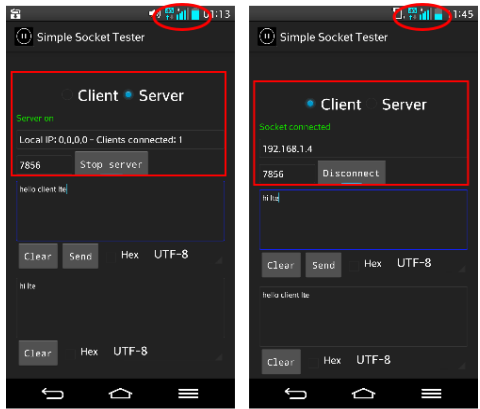

(b) TCP session between 2 UEs

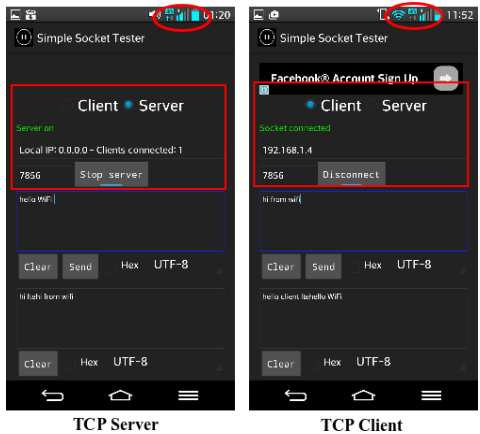

(c) Seamless offload of TCP client to WiFi network

Fig. 16: Experiments with two UE mobiles.

[15] A. Rahmati, C. Shepard, C. Tossell, L. Zhong, P. Kortum, A. Nicoara, and J. Singh, "Seamless TCP Migration on Smartphones without Network Support," *IEEE Transactions on Mobile Computing*, vol. 13, no. 3, pp. 678–692, Mar. 2014.

[16] S. Nirjon, A. Nicoara, C.-H. Hsu, J. Singh, and J. Stankovic, "MultiNets: Policy Oriented Real-Time Switching of Wireless Interfaces on Mobile Devices," in *18th IEEE Symposium on Real-Time and Embedded Technology and Applications (RTAS)*, Apr. 2012, pp. 251–260.

[17] "A 3G/LTE Wi-Fi Offload Framework: Connectivity Engine (CnE) to Manage Inter-System Radio Connections and Applications," https://www.qualcomm.com/documents/3g-lte-wifi-offload-framework, Dec. 2015.

[18] "3GPP TS 24.312 : Access Network Discovery and Selection Function (ANDSF) Management Object (MO), (Rel. 10)," http://www.3gpp.org/DynaReport/24312.htm, Dec. 2015.

[19] P. Loureiro, M. Liebsch, and S. Schmid, "Policy routing architecture for IP flow mobility in 3GPP's Evolved Packet Core," in *IEEE GLOBECOM Workshops*, Dec. 2010, pp. 2000–2005.

[20] Q. Cui, Y. Shi, X. Tao, P. Zhang, R. Liu, N. Chen, J. Hamalainen, and A. Dowhuszko, "A unified protocol stack solution for LTE and WLAN in future mobile converged networks," *IEEE Wireless Communications*, vol. 21, no. 6, pp. 24–33, Dec. 2014.

[21] M. Yang, Y. Li, D. Jin, L. Su, S. Ma, and L. Zeng, "OpenRAN: a software-defined ran architecture via virtualization," in *Proceedings of the ACM SIGCOMM*, Aug. 2013, pp. 549–550.

[22] S. H. Yeganeh, A. Tootoonchian, and Y. Ganjali, "On scalability of software-defined networking," *IEEE Communications Magazine*, vol. 51, no. 2, pp. 136–141, February 2013.

[23] "Host Model," http://en.wikipedia.org/wiki/Host_model, Dec. 2015.

[24] T. Melia and S. Gundavelli, "Logical Interface Support for multi-mode IP Hosts," http://tools.ietf.org/html/draft-ietf-netext-logical-interface-support-09, Dec. 2015.

[25] J. Lee, "Dmm wg draft: Pmipv6-based distributed mobility management," https://tools.ietf.org/html/draft-jaehwoon-dmm-pmipv6-01, Jun. 2013.

[26] "NS-3," http://www.nsnam.org/documentation/, Dec. 2015.

[27] "OpenLTE," http://openlte.sourceforge.net, Dec. 2015.

[28] "GNU radio," https://en.wikipedia.org/wiki/GNU_Radio, Dec. 2015.

[29] "ETTUS SDR," http://www.ettus.com/product/category/USRP-Bus-Series, Dec. 2015.

[30] "hostapd," https://wireless.wiki.kernel.org/en/users/documentation/hostapd, Dec. 2015.

[31] "dnsmasq," http://www.thekelleys.org.uk/dnsmasq/doc.html, Dec. 2015.

[32] "3GPP TS 24.234 : 3GPP system to Wireless Local Area Network (WLAN) interworking; WLAN User Equipment (WLAN UE) to network protocols, (Rel. 10)," http://www.3gpp.org/DynaReport/24234.htm, Dec. 2015.